\documentclass[aps, prd, reprint,showpacs,  preprintnumbers,amsmath,amssymb,superscriptaddress,  nofootinbib]{revtex4-1}

\usepackage{amssymb, amsmath, braket, nicefrac, graphicx} 
\usepackage[tight]{subfigure}
\usepackage{relsize}
\usepackage{lineno}
\usepackage{setspace}
\usepackage{ esint }
\usepackage{chngpage}
\usepackage{epsfig}
\usepackage{multirow}
\usepackage{rotating}
\usepackage{float}
\usepackage{color}
\usepackage{xcolor}
\usepackage{appendix}
\usepackage{ amstext }
\usepackage{ gensymb }
\usepackage{ textcomp }
\usepackage{ wasysym }

\usepackage{upgreek}

\usepackage{pstricks}

\newcommand{\numu}{\mbox{$\nu_{\mu}$}}                   % nu_mu 
\newcommand{\nutau}{\mbox{$\nu_{\tau}$}}                 % nu_tau 

\newcommand{\anue}{\ensuremath{\bar{\nu}_{e}}}

\newcommand{\anumu}{\ensuremath{\bar{\nu}_{\mu}}}

\newcommand{\simgt}{\,\hbox{\lower0.6ex\hbox{$\sim$}\llap{\raise0.6ex\hbox{$>$}}}\,}
\newcommand{\simlt}{\,\hbox{\lower0.6ex\hbox{$\sim$}\llap{\raise0.6ex\hbox{$<$}}}\,}

\definecolor{maroon}{RGB}{162,10,10}

\usepackage{hyperref}
\usepackage[all]{hypcap}
\hypersetup{
    colorlinks,
    citecolor=black,
    filecolor=black,
    linkcolor=black,
    urlcolor=black
}

\makeatletter

\renewenvironment{figure}
  {\def\@captype{figure}}
  {}
\makeatother
\begin{document}

\title{\bf  Nonmaximal $\theta_{23}$ mixing at NOvA from neutrino decoherence}

\newcommand{\Tufts}{Physics Department, Tufts University, Medford, Massachusetts 02155, USA}
\newcommand{\APC}{APC, Universit\'{e} Paris Diderot, CNRS/IN2P3, Sorbonne Paris Cit\'{e}, F-75205 Paris, France}

\author{Jo\~{a}o A. B. Coelho}\affiliation{\Tufts} \affiliation{\APC}
\author{W. Anthony Mann}\affiliation{\Tufts}
\author{Saqib S. Bashar}\affiliation{\Tufts}


\begin{abstract}

In study of muon neutrino disappearance at 810 km, the NOvA experiment finds flavor mixing
of the atmospheric sector to deviate from maximal ($\sin^2\theta_{23} = 0.5$) by 2.6\,$\sigma$.  
The result is in tension with the 295-km baseline measurements of T2K 
which are consistent with maximal mixing. We propose that $\theta_{23}$ 
is in fact maximal, and that the disagreement is harbinger of environmentally-induced decoherence. 
The departure from maximal mixing can be accounted for by an energy-independent decoherence 
of strength $\Gamma = (2.3 \pm 1.1) \times 10^{-23}$ GeV.

\end{abstract}

\maketitle

%\section{Introduction}
For nearly two decades, experimental investigations of atmospheric and accelerator-beam $\numu$ and $\anumu$ neutrinos
have indicated the $\theta_{23}$ mixing angle to be compatible with $45^{o}$ (i.e. $\sin^2\theta_{23} = 0.5$), implying the
$\nu_3$ mass eigenstate to be comprised of $\numu$ and $\nutau$ flavors in nearly equal amounts.
This trend has been taken as evidence for a $\mu - \tau$ flavor symmetry that may underwrite the pattern exhibited by the 
3 x 3 Pontecorvo-Maki-Nakagawa-Sakata lepton-flavor mixing matrix~\cite{Xing-Zhao}.    At present, experimental analyses of 
muon neutrino disappearance by T2K, MINOS, Super-Kamiokande, and IceCube continue to report $\sin^2\theta_{23}$ allowed
regions that are consistent with maximal mixing -- see for example Ref.~\cite{Smirnov-Talk}.
The new measurement reported by NOvA however, breaks the trend.   
Using a 14 kton detector equivalent exposure of 6.05$\times 10^{20}$
protons on target, NOvA determines two statistically degenerate values 
for $\sin^2\theta_{23}$ in the normal mass hierarchy (NH) at 68\% confidence level:
$0.404^{ +0.030}_{-0.022}$  and  $0.624^{ +0.022}_{-0.030}$.   The results indicate departure, 
at 2.6 $\sigma$ significance, from maximal $\theta_{23}$ mixing~\cite{NOvA-nonmax}.

That a flavor symmetry may be operative and partially broken
is a tantalizing situation, for the amount of symmetry
breaking is potentially informative about underlying structures.  
On the other hand it is not particularly shocking that a current
experiment might reveal a departure from maximal mixing,
since it is often the case that symmetries are inexact at some level.
More disconcerting is that the NOvA result diverges from 
the findings of the other state-of-the-art experiment that uses
a fixed long baseline, namely T2K.   Understandably, the tension 
between these measurements has drawn the attention of proponents of exotic
oscillation effects.  An obvious difference between the experiments 
is their baselines; the T2K far detector is located at 295 kilometers
from the accelerator beam site, while the NOvA baseline is 810 kilometers.    
For conventional 3-flavor vacuum oscillations, muon neutrino
disappearance depends on $L/E_{\nu}$ and the baseline difference 
should be of little consequence.   However for exotic physics effects 
promoted by propagation through matter or merely by extended propagation 
through space-time, the difference in the baselines can be relevant.

%\subsection{Alternatives to non-maximal mixing}
Proponents of exotic oscillation effects take the view that 
there may be an exact $\mu - \tau$ flavor symmetry.   
The NOvA result is then to be understood as a harbinger of new physics.   
One possibility is that neutrinos propagating through
the Earth's crust are subjected to non-standard interactions (NSI).    
The tension between the NOvA and T2K measurements
of the $\theta_{23}$ mixing angle received treatment 
in two recent NSI analyses~\cite{Yasuda-2016, Liao-2016}.  
The NSI scenarios developed by these works are  
quite different;  each analysis invokes a different set of sizable 
NSI couplings, and some of the couplings are required to have strengths 
comparable to the Mikheyev-Smirnov-Wolfenstein (MSW) matter effect~\cite{ref:MSW}. 

\smallskip

It is proposed in this work that a simple model of neutrino decoherence driven by weak 
coupling to a dissipative environment
offers another way wherein $\mu - \tau$ flavor symmetry remains exact 
while the disagreement between the NOvA and T2K 
$\numu$-flavor disappearance results is explained.    
While full-bore decoherence models were `run out of town' a decade ago, 
overwhelmed by the accumulation of data that showed oscillations to be
the dominant effect, the possibility remains that propagating neutrinos 
may decohere very gradually as they oscillate.    Such behavior
is observed in a variety of quantum systems that 
are `open' to environmental influences, and the phenomenology for 
describing these systems is well-developed.   For evolving neutrino states, the pervasive environment
might introduce new physics originating beyond the standard electroweak model, e.g. perturbations
arising from spacetime itself and its Planck-scale dynamics~\cite{Spacetime-1995, Lisi-2000, Barenboim-2006}.   

Environmentally-induced neutrino decoherence is to be distinguished from neutrino wave packet decoherence.
The latter is a quantum wave effect that one may expect to occur based on known physics -- no beyond-the-Standard-Model 
mechanism is needed.   Neutrino wave packets have received abundant
treatment from both relativistic quantum-mechanical and quantum-field theoretic perspectives~\cite{Chan-2016}.  
In recent times the disappearance of reactor $\anue$ at nine different baselines of the Daya Bay experiment was examined 
for wave-packet effects, however none were found~\cite{Daya-Bay-2016}.   
Neglecting higher-order dispersion-effect terms, there is general agreement concerning the basic form that neutrino wave-packet
decoherence would take~\cite{Giunti-1998, Chan-2016}.
For the $\numu$ survival-oscillation probability, the prerequisite integration over momentum space 
and averaging over the time from production to detection 
introduces exponential damping factors that multiply each of the oscillatory terms.
The damping factors have the form $\exp\{ - (L/L^{coh}_{ij})^2\}$, where $L$ is the baseline length, $ij =$ 21 or 31 or 32, and 
$L^{coh}_{ij}$ is the coherence length: $L^{coh}_{ij} =  (4\sqrt{2} \sigma_{x} E_{\nu}^2)/ |\Delta m^2_{ij}|$.   Importantly,
the coherence length is proportional to the width $\sigma_x$ of the mass eigenstate wave packets in coordinate space and
to the square of the neutrino energy, $E_{\nu}$.   This means that the exponential damping depends strongly on 
neutrino energy as well as on baseline:  $ (L/L^{coh}_{ij})^2 = (|\Delta m^2_{ij}|^2 \,L^2)/(32\, \sigma_{x}^2 \, E_{\nu}^4)$.
For accelerator-based long baseline experiments such as T2K and NOvA, the processes at neutrino production at 
detection are nearly the same, hence $\sigma_x$ should be of similar magnitude.
Moreover, in long baseline experiments the oscillation phase $\phi = (\Delta m^2_{32} L)/4 E_{\nu}$ 
is chosen near unity, consequently $ (L/L^{coh}_{ij})^2 \propto \phi^4/L^2$ is decreasing with baseline.
Thus wave packet decoherence
is not viable as an effect that could account for the emergence 
of apparent non-maximal mixing with a longer baseline.
On the other hand, as will be elaborated,
an environmentally-driven decoherence that depends only
on path length can account for the emergence 
of apparent non-maximal mixing at longer oscillation baselines.
Furthermore the strength of the decoherence 
required to do this is not contradicted by the 
upward-going muon data of Super-Kamiokande~\cite{Lisi-2000, Fogli-2007}.

%\section{Neutrino evolution in a dissipative environment}
The survival probability for $\numu$ flavor neutrinos propagating 
through vacuum, $\numu \rightarrow \numu$, is approximately described by 2-flavor mixing: $ \mathcal{P}_{\mu\mu}
= ~ 1 - \sin^{2}2\theta_{23} \cdot \sin^{2}\phi$. 
A tacit assumption is that propagating neutrinos comprise a closed quantum system.
Most systems, however, are inherently open to an environment and are
potentially susceptible to dissipative interactions with it.  
The dissipative effect considered here is a decoherence effect that acts on the quantum interference and 
damps the oscillating terms in the neutrino oscillation probabilities.

A phenomenology that allows for dissipative interactions with an environment is provided by a
density matrix formalism and the quantum analogue 
of Liouville's theorem in classical statistical mechanics~\cite{Gorini-1978, Lindblad}.  
In particular, the Lindblad master equation is generally regarded as an appropriate framework for 
investigating neutrino decoherence~\cite{Lisi-2000, Ohlsson-2001, Benatti-2001, Gago-2001, 
Barenboim-2006, Fogli-2007, Oliveira-EurPhys-2010, Oliveira-2013, Oliveira-2014, Guzzo-2016}.     
The presence of weakly perturbative dynamics is parameterized by an added ``dissipator" term,
%%%%%%%%%%%%%%%%%%%%%%%%%%%%%%%%%%%
\begin{equation} \label{eq:time-evolution-2}
    \frac{d}{dt}\hat{\rho}_{\nu}(t) = - i [\hat{H} , \hat{\rho}_{\nu}(t) ] - \mathcal{D}[\hat{\rho}_{\nu}(t)].
  \end{equation}
%%%%%%%%%%%%%%%%%%%%%%%%%%%%%%%%%%
The dissipation term $\mathcal{D}[\hat{\rho}_{\nu}(t)]$ is constructed using a set of $N^2 -1$ operators, 
 $\hat{D}_{n}$, where $N$ is the dimension of the 
 Hilbert space of interest (so $N = 2$ for two-flavor oscillations and the $\hat{D}_{n}$ 
 are linear combinations of the Pauli spinors plus the unit matrix).    
 Constraints can be placed on the $\hat{D}_{n}$ arising from mathematical considerations 
 and from the laws of thermodynamics.   For example, it may be assumed that the von Neumann entropy, 
 $S = {Tr} (\hat{\rho}_{\nu} \ln\hat{\rho}_{\nu} )$, increases with time and this is enforced 
 by requiring the $\hat{D}_{n}$ to be hermitian.  In addition,
 conservation of the average value of the energy, calculated as ${Tr}(\hat{\rho}_{\nu} \hat{H})$, 
 can be assured by requiring the $\hat{D}_{n}$ to commute with $\hat{H}$.  
 For two-flavor mixing describing vacuum oscillations of the atmospheric sector,
 the phenomenology is reducible 
 to a form in which decoherence is promoted by a single exponential
 damping term containing one free parameter, $\Gamma_{32}$.    
 The probability for $\numu$ disappearance oscillations, obtained by tracing 
the $\ket{\numu}$ state projector (expressed in mass basis) 
over the time-evolved density matrix~\cite{Lisi-2000, Oliveira-EurPhys-2010}, is 
 %%%%%%%%%%%%%%%%%%%%%%%%%%%%%%%%%%%%%%
\begin{equation} \label{eq:trace-decohere}
\mathcal{P}_{\mu\mu}^{(2\nu)} = 1 - \frac{1}{2} \sin^{2}2\theta_{23} \cdot \left[1  - e^{-\Gamma_{32} L} \cdot \cos( \frac{ \Delta m^2_{32}}{2 E_{\nu}}L )\right].
 \end{equation}
 %%%%%%%%%%%%%%%%%%%%%%%%%%%%%%%%%%%%%%%
Equation \eqref{eq:trace-decohere} has a resemblance 
to expectations for $\numu$ survival in the presence of neutrino decay~\cite{Bertlmann-2006, Garcia-2008}, however
there are differences.   For oscillations with decay, the decay rate gives the damping constant 
and, due to the Lorentz boost, the damping carries an $E_{\nu}^{-1}$ 
dependence.   Moreover neutrino decay models lead to damping of constant terms as well as oscillatory terms in the survival probability,
while damping from decoherence is limited to oscillatory terms.

%\subsection{Constant versus energy power-law coupling}
%\label{sec:power-law-discuss}
The interaction of neutrinos with their environment need not be constant 
-- it could depend upon $E_{\nu}$.   Previous investigations
of neutrino decoherence models explored this possibility using integer power-law forms 
for the decoherence parameter~\cite{Lisi-2000, Fogli-2007, Oliveira-2014}:
$\Gamma_{32} = \Gamma_{0} \cdot (\frac{E_{\nu}}{[GeV]})^{n}.$
In the absence of a model for environmental influence, many researchers 
have focussed on the n = 0 case, however power-law forms with $ n = 0, \pm 1, \pm 2$ have been regarded as possibilities.
The case n = 2 is strongly constrained by the Super-Kamiokande atmospheric data~\cite{Lisi-2000}; these constraints become
weaker with a slower rate of energy increase or with a decreasing energy dependence.   For neutrino
mixing in the solar sector, the decoherence parameter $\Gamma_{21}$ with n = -1 (and presumably n = -2) is strongly constrained
by the solar plus KamLAND data~\cite{Fogli-2007}.    In any event the negative integer power-law forms do not work for the
scenario considered here.  This leaves n = 0 as the simplest choice for the scenario proposed.    Support for
this choice is given by the decoherence model fit results 
of Oliveira {\it et al.}~\cite{Oliveira-2014} to the $\numu$ and $\anumu$ 
disappearance oscillation data of the MINOS experiment at 735 kilometer baseline.   
For the n = 0 power law, their best-fit with conventional 2-flavor oscillations 
gives $\sin^{2}(2\theta_{23}) = 0.92\,^{+0.06}_{-0.07}$, 
while the decoherence model yields $\sin^{2}(2\theta_{23}) = 0.98\,_{-0.08}$ with
$\Gamma_{32} = 3.10\,^{+2.37}_{-2.49} \times 10^{-23}$ GeV.

%\subsection{Extension to $3\nu$ oscillations with matter effect}
The NOvA measurement is based on data analysis using $3\nu$ oscillations 
with the MSW matter effect and so, for an accurate evaluation
of decoherence, it is necessary to extend the phenomenology 
to a comparable framework.   
The Hermitian operators of the Lindblad equation, namely $\hat{H}, \hat{\rho}_{\nu}(t)$, and the eight $\hat{D}_{n}$ operators, can be
expanded in terms of the Gell-Mann SU(3) basis matrices and the $3 \times3 $ unit matrix.  This enables a re-formulation of the evolution equation
that includes a $8 \times 8$ matrix of parameters, $\mathcal{D}_{kl}$~\cite{Gago-2002, Oliveira-2016}.    The requirement $[ \hat{H}, \hat{D}_{n} ] = 0$
constrains the $\mathcal{D}_{kl}$ matrix to be diagonal, with elements involving 
only three positive, real-valued parameters: $\Gamma_{21}, \Gamma_{31}$, and $\Gamma_{32}$~\cite{Oliveira-2016}.
The time evolution of the density matrix for
three-neutrino oscillations in vacuum has been solved and the 
oscillation probabilities with inclusion of decoherence obtained 
(see Eqs.~(2.4) and (2.6) in Ref.~\cite{Farzan-2008}).   
Here we proceed by replacing the mixing angles 
and mass-splittings with their corresponding matter effective values:
%%%%%%%%%%%%%%%%%%%%%%%%%%%%%%%%%%%%%%%%%%
\begin{equation} 
\nonumber
\label{General-survival}
\mathcal{P}_{\mu\mu}^{(3\nu)} = 1 - 2 \,\sum_{ j > k} \{ |\tilde{U}_{\mu j}|^{2}|\tilde{U}_{\mu k}|^{2} (1 -  e^{-\Gamma_{jk}L} \cos \tilde\Delta_{j k} L ) \} ,                                            
\end{equation}
%%%%%%%%%%%%%%%%%%%%%%%%%%%%%%%%%%%%%%%%%%%
where $\tilde{U}_{\mu i}$ are elements of an effective mixing matrix, 
and $\tilde\Delta_{j k}$ are the mass-splitting forms ($\Delta m^2_{jk}/2 E_{\nu}$)
augmented by factors arising from matter effects.

A very good approximation for 
$\mathcal{P}_{\mu\mu}^{(3\nu)}$ with matter effects (sans decoherence) is presented 
in Ref.~\cite{Choubey-2006}, obtained under the assumption that $\Delta m^{2}_{21}$ = 0.   
This approximate form (see Eq. (27) of~\cite{Choubey-2006}) can be  
rearranged to allow the decoherence factors of Eq.~\eqref{General-survival} to be included, yielding
\begin{equation} \label{good-approximation-1}
\begin{split}
\mathcal{P}_{\mu\mu}^{(3\nu)} \approx 1 
&- \frac{1}{2} \sin^2\tilde{\theta}_{13} \sin^2 2\theta_{23} \left[ 1 - e^{-\Gamma_{21} L} \cdot \cos 2 \tilde{\phi}_{-} \right] \\
&- \frac{1}{2} \cos^2\tilde{\theta}_{13} \sin^2 2\theta_{23}  \left[ 1 - e^{-\Gamma_{32} L} \cdot \cos 2 \tilde{\phi}_{+} \right] \\
&- \frac{1}{2} \sin^{2} 2\tilde{\theta}_{13} \sin^4 \theta_{23} \left[ 1 -  e^{-\Gamma_{31} L} \cdot \cos 2 \tilde{\phi}_{0} \right],
\end{split}
\end{equation}
where
\begin{equation} \label{good-approx-3a}
\tilde{\phi}_{0} \equiv  \phi \cdot \sqrt{ (\cos 2\theta_{13} - \hat{A})^2 + \sin^2 2\theta_{13}} ~, 
\end{equation}
\begin{equation} \label{good-approx-3b}
\tilde{\phi}_{\pm} \equiv \frac{1}{2} \left[ (1 + \hat{A}) \cdot \phi\, \pm\,  \tilde{\phi}_{0} \right], 
\end{equation}
with the matter potential $\hat{A} = (2\sqrt{2} G_{F} N_{e} E_{\nu})/\Delta m^2_{31}$ 
and with $\tan 2 \tilde{\theta}_{13} = \sin 2\theta_{13}/(\cos 2\theta_{13} - \hat{A})$.

\smallskip

In Eq.~\eqref{good-approximation-1}, the first term on the right-hand side 
is affected by the $\Gamma_{21}$ decoherence parameter.   
However that parameter is assumed to be negligible here, motivated by
fits to the available solar plus KamLAND neutrino data~\cite{Fogli-2007} which obtained 
for the same n=0 power-law form:  $\Gamma_{21} < 0.67 \times 10^{-24}\, \text{GeV}$ at $95\% ~\text{C.L}$. 
The three $\Gamma_{ij}$ parameters of the $\mathcal{D}_{kl}$ matrix are related
by the requirement of complete positivity~\cite{Lindblad, Benatti-1997} in such a way that if $\Gamma_{21}$ = 0, then
$\Gamma_{32} = \Gamma_{31} \equiv \Gamma $\,(see Ref.~\cite{Oliveira-2016}, Sec.~2).  
With Eq.~\eqref{good-approximation-1} it is readily seen that 
in the limit of vacuum oscillations and $\theta_{13} \rightarrow 0$, the first and third terms 
go to zero and the second term goes over to Eq.~\eqref{eq:trace-decohere}.

%\section{Determining $\Gamma$ from apparent nonmaximal $\theta_{23}$}
In order to measure the $\theta_{23}$ mixing angle, experiments look 
for the oscillation probability in the vicinity of the first oscillation minimum.
For $\numu$ flavor disappearance in Standard Model (SM) oscillations, the survival minima
in the presence of non-maximal $\theta_{23}$ mixing will be shifted from null probability to a small probability.
This effect, and the reinterpretation that is possible for it, 
are readily discerned in the vacuum, $2\nu$-oscillation formulas.   
In the case of no decoherence, the survival probability indicates that 
non-maximal $\theta_{23}$ can give an upward shift of ($1.0 - \sin^{2}2\theta_{23}^{SM}$).   Equation~\eqref{eq:trace-decohere}
indicates that the same probability shift can arise with maximal $\theta_{23}$ if decoherence is operative;  an upwards shift 
of (1 - $\frac{1}{2} (1 + e^{-\Gamma_{32}L}))$ is to be expected.   

These same trends are predicted by the $3\nu$ oscillation formula 
of Eq.~\eqref{good-approximation-1}, which provides a more accurate venue for relating
the decoherence parameter, $\Gamma$, to measurements of 
apparent non-maximal $\theta_{23}$ mixing.   In long-baseline experiments, the matter potential
$\hat{A}$ has a small value close to the oscillation minimum: 
$\hat{A}^{(min)} \sim (0.085 \, E^{(min)}_{\nu}$/GeV).    
Consequently the oscillation minimum for $\mathcal{P}_{\mu\mu}^{(3\nu)}$
of Eq.~\eqref{good-approximation-1} lies very close to the minimum for vacuum oscillations,
$E^{(min)}_{\nu} = (\Delta{m^2_{31}}L)/{2\pi}$.

The survival probability at the oscillation minimum, $\mathcal{P}_{\mu\mu}^{(min)}$, 
can then be estimated by expanding Eq.~\eqref{good-approximation-1}
about $E^{(min)}_{\nu}$ and retaining terms with any product 
of $\hat{A}^{(min)}$ and $\sin \theta_{13}$ up to third order ($\sim0.3\%$).
This procedure yields
%%%%%%%%%%%%%%%%%%%%%%%%%%%%%%%
\begin{equation}
%\nonumber
\label{eq:approx-P-min}
\begin{split}
&P_{\mu\mu}^{(min)} \approx  1 - (1+e^{-\Gamma L}) \times  \\
&\left[\frac{1}{2}\sin^22\theta_{23}  - 2\sin^2\theta_{23}\cos2\theta_{23}\sin^2\theta_{13}(1 + 2\hat{A}^{(min)})\right].
\end{split}
\end{equation}
%%%%%%%%%%%%%%%%%%%%%%%%%%%%%%%
The value of $\Gamma$ predicted by the decoherence scenario can be found by equating the probabilities
\begin{equation}
%\nonumber
P_{\mu\mu}^{(min)}(\Gamma=0, \sin^2\theta_{23}^{SM})=P_{\mu\mu}^{(min)}(\Gamma, \sin^2\theta_{23}=0.5).
\end{equation}
The result is:
\begin{equation}
%\nonumber
\label{eq:Gamma-result-1}
\begin{split}
&e^{-\Gamma L} = 8 \cdot \sin^2\theta_{23}^{SM} \times \\
&\left[1 - \sin^2\theta_{23}^{SM} - (1 + 2\hat{A}^{(min)})\sin^2\theta_{13}\cos2\theta_{23}^{SM}\right] - 1
\end{split}
\end{equation}
where $\hat{A}^{(min)} \simeq L/5800$ km.    Evaluating the above equation 
at  810-km baseline using NOvA's reported NH 
values, $\sin^2\theta_{23}^{SM}=\{ 0.404^{ +0.030}_{-0.022}$  \mbox{ or }  $0.624^{ +0.022}_{-0.030} \}$,  
together with $\sin^2\theta_{13}=0.0219$, we obtain:
\begin{equation}
\label{result-NOvA-baseline}
\Gamma = (2.3 \pm 1.1 ) \times 10^{-23} \, \text{GeV}.
\end{equation}
Note that the value for the decoherence parameter is the same for
either of the octant solutions for $\sin^2\theta_{23}^{SM}$.    Our result
is compatible with the limits reported in Refs.~\cite{Lisi-2000, Fogli-2007}, based upon 
comparison of Super-Kamiokande lepton distributions in zenith angle with predictions for muon survival using
a constant (n = 0) decoherence parameter: $\Gamma_{32} < 4.1\,(5.5) \times 10^{-23}$ GeV at 95\%\,(99\%) C.L.

\begin{figure}%%[!htb]
\begin{adjustwidth}{1.0in}{1.0in}
\centering
\resizebox{\columnwidth}{!}{\includegraphics[angle=0, trim = 0mm 0mm 0mm 0mm, clip]{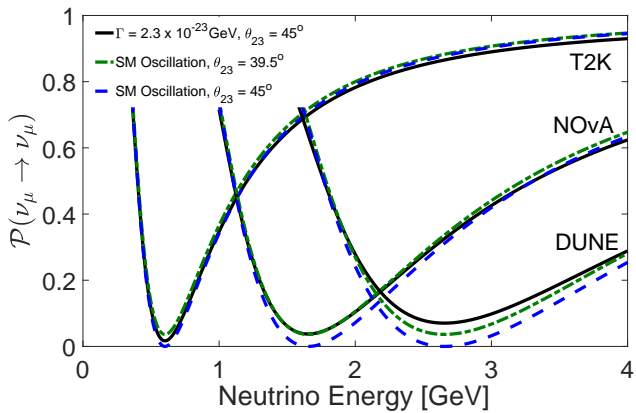}}
\end{adjustwidth}
\caption{Muon-neutrino survival versus $E_{\nu}$ at the T2K, NOvA, and DUNE baselines in the vicinity of their respective first minima.
The probability distributions compare standard oscillations 
with maximal and non-maximal $\theta_{23}$ mixing (long-dash, short-dash curves), 
to oscillations with maximal mixing plus decoherence (solid-line curve).}
\label{fig:Fig01}
%\end{center}
\end{figure}

%\section{Decoherence effect in $\theta_{23}$ measurements}
Figure~\ref{fig:Fig01} shows $\numu$ survival probabilities for the 
ongoing experiments T2K and NOvA, and for the 1300-kilometer baseline of DUNE.    For each baseline, the survival probability
versus $E_{\nu}$ is displayed {\it i)} for standard oscillations with maximal mixing (long-dash curve),  {\it ii)} for standard oscillations with 
NOvA non-maximal mixing (short-dash curve), and {\it iii)} for maximal mixing oscillations with decoherence (solid-line curve).   At the first
minimum for the NOvA baseline, non-maximal mixing is indistinguishable (by construction) from maximal mixing plus decoherence.  At the 
first minimum for T2K, the probability curves indicate that the decoherence scenario is more difficult to distinguish from maximal 
mixing than is NOvA nonmaximal mixing.   Thus if decoherence rather than non-maximal mixing is operative, this situation may explain
the apparent tension between NOvA and T2K measurements of $\numu$-flavor survival oscillations.
 The probability curves at the first minimum of DUNE show that distinguishing among 
predictions of the three hypotheses becomes easier at longer baselines.

%%%%%%%%%%%%%%%%%%%%%%%%%%%%%%%%%%%%%%%%%%%  Figure 01
%\begin{figure}
%\begin{center}
%\includegraphics[width=8.8cm]{Fig01.pdf}
%%%%%%%%%%%%%%%%%%%%%%%%%%%%%%%%%%%%%%%%%%%%%%

\begin{figure}%%[!htb]
\begin{adjustwidth}{0in}{0in}
\centering
\resizebox{\columnwidth}{!}{\includegraphics[angle=0, trim = 0mm 0mm 0mm 0mm, clip]{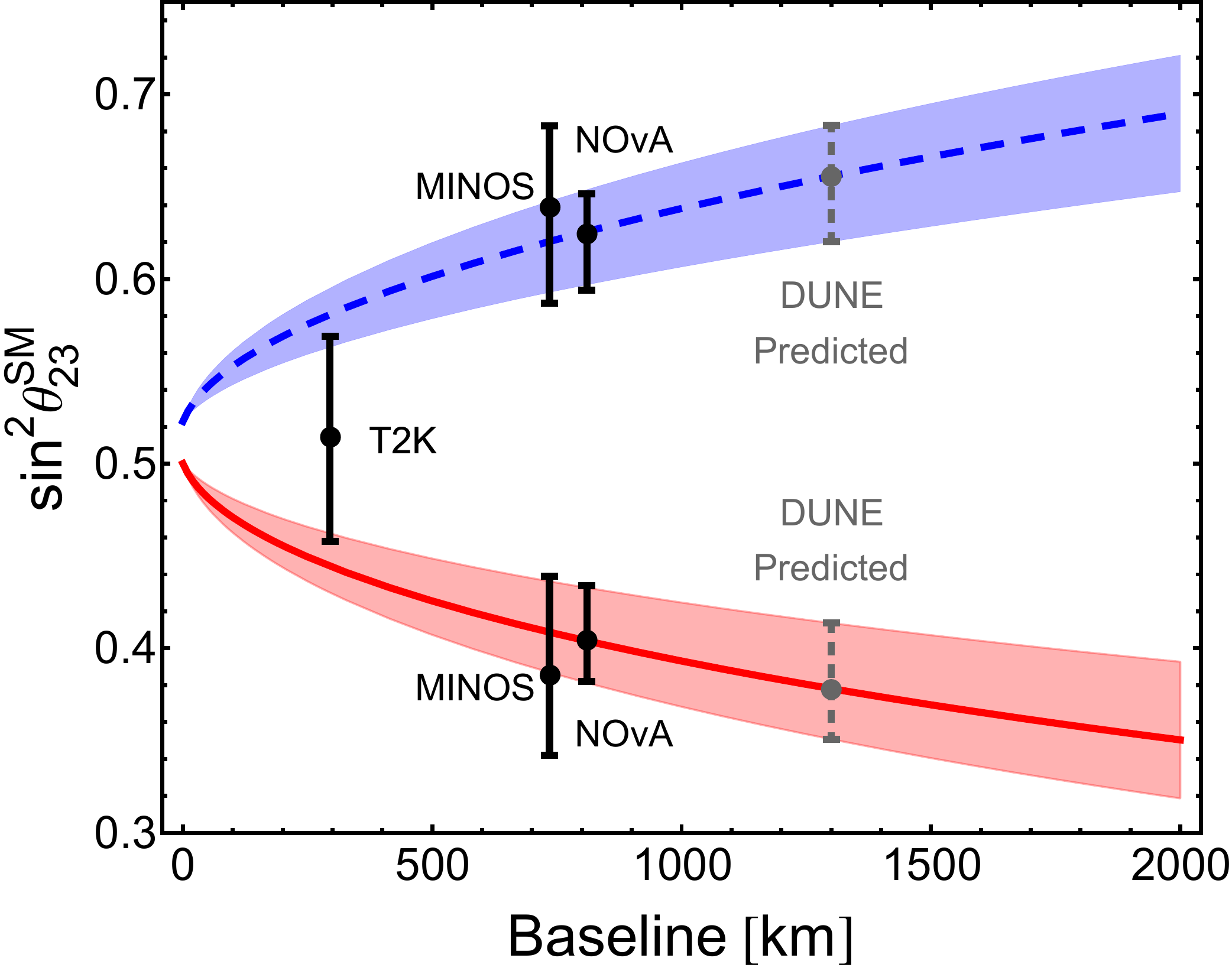}}
\end{adjustwidth}
\caption{Depiction of trend to larger apparent deviation from maximal $\theta_{23}$ mixing with increasing 
oscillation baseline, as predicted by the maximal mixing plus decoherence oscillation scenario.   The
predictions are double-valued and asymmetric about $\sin^{2}\theta_{23}$ = 0.5, 
reflecting the octant degeneracy inherent to measurement of $\theta_{23}$ in the presence of matter effects.
The curves displayed are for the normal mass hierarchy.}
\label{fig:Fig02}
%\end{center}
\end{figure}

Having determined the $\Gamma$-parameter value (Eq.~\eqref{result-NOvA-baseline}), it becomes possible to extract
from Eq.~\eqref{eq:Gamma-result-1} the values for $\sin^2\theta_{23}^{SM}$ to be expected at any long baseline from a
conventional SM oscillation analysis that ignores decoherence.   Upon inverting Eq.~\eqref{eq:Gamma-result-1} for this
purpose, one obtains two solutions for $\sin^2\theta_{23}^{SM}$ at each baseline, one for each of the octants.
The solutions as a function of increasing baseline follow two distinct trajectories as is shown by 
Fig.~\ref{fig:Fig02}, with the apparent $\sin^2\theta_{23}^{SM}$ value for each solution displayed on the y-axis.
The solution points at each baseline are not symmetrically located about $\sin^2\theta_{23}^{SM} = 0.5$, reflecting the 
interplay of 3-$\nu$ mixing with matter effects.     Included in the Figure are experimental data points and errors
representing beam-only $\numu$ disappearance results reported 
by NOvA~\cite{NOvA-nonmax}, MINOS~\cite{MINOS-2013}, and T2K~\cite{T2K-2014}.

 Figure~\ref{fig:Fig02} shows that long baseline experiments using a conventional SM analysis will, at shorter baselines, 
tend to infer that $\theta_{23}$ mixing lies close to maximal  -- assuming that the mixing is indeed maximal.
At the T2K baseline, a precise measurement could in principle discern a deviation from maximal, 
$\Delta \theta_{23} \simeq \pm 4^{\circ}$,  although in practice a deviation of this size would be difficult to observe.   
At MINOS+ and NOvA however, $\Delta \theta_{23}$ grows to $\simeq \pm 6^{\circ}$, 
and so an apparent deviation from maximal mixing is more readily obtained.   At the baseline of DUNE, a 
conventional oscillations measurement is predicted to report a larger excursion of $\theta_{23}$ 
from maximal mixing than that deduced at NOvA: $\Delta \theta_{23} \simeq \pm 8^{\circ}$.

%\smallskip
%%%%%%%%%%%%%%%%%%%%%%%%%%%%%%%%%%%%%%%%%%%%%%%  Figure 02
%\begin{figure}
%\begin{center}
%\includegraphics[width=8.8cm]{Fig02.pdf}
%%%%%%%%%%%%%%%%%%%%%%%%%%%%%%%%%%%%%%%%%%%%%% 
\smallskip

The decoherence parameter value presented in Eq.~\eqref{result-NOvA-baseline}
is inferred from the values reported by NOvA. 
An alternative approach could be based on a combined fit to the three long baseline experiments. 
As an exploratory trial, a chi-square fit for the decoherence parameter was carried out 
assuming errors are Gaussian in the value of the probability minimum
of Eq.~\eqref{eq:approx-P-min}, using the data points and errors displayed in Fig.~\ref{fig:Fig02}.  
This fit ($\chi^{2}/d.o.f. = 1.02$) yields a somewhat lower value for the decoherence parameter:
$\Gamma = (1.8 \pm 0.9) \times 10^{-23}$ GeV.  The central trajectories are very similar to those displayed in Fig.~\ref{fig:Fig02}
but lie slightly closer to the horizontal centerline at $\sin^{2}\theta^{SM}_{23}$; they fall at the $1\,\sigma$ limits of T2K and remain 
well within the $1\,\sigma$ ranges of MINOS and NOvA.   A more incisive treatment
requires attention to details of the experiments and is not pursued here.
Further insights might be gleaned by considering the high energy $\numu$ event samples 
($E_{\nu} > 4$ GeV) that are accessible to MINOS+ and DUNE.

\smallskip

To summarize:  It is proposed that a small decoherence effect whose strength lies
just below the current upper limit can account for the non-maximal mixing observation at NOvA while indicating why $\theta_{23}$
appears to be more nearly maximal at T2K.  The decoherence effect is characterized by a single, 
energy-independent damping parameter, $\Gamma = (2.3 \pm 1.1 ) \times 10^{-23} \, \text{GeV}$.

 If maximal $\theta_{23}$ mixing plus decoherence are operative, but $\theta_{23}$ measurements
continue to be expressed in terms of standard oscillations without decoherence, then certain trends in neutrino results 
can be anticipated: {\it (i)}  NOvA will continue  to report non-maximal mixing.   {\it (ii)} T2K results will gradually shift 
from maximal to nonmaximal $\theta_{23}$ mixing but with a deviation from maximal that is always less than that reported by NOvA.
{\it (iii)} This apparent tension will hold regardless of whether $\numu$ and $\anumu$ data samples are treated separately or together at each of the baselines.
{\it (iv)}  At DUNE, a larger (apparent) deviation from maximal $\theta_{23}$ mixing will be observed than that reported by NOvA.
Specifically, $\numu$ disappearance in DUNE will appear to be governed by mixing at  
strength $\sin^{2}\theta_{23} \simeq 0.38$ for the lower octant NH solution. ÊSensitive tests of decoherence using atmospheric neutrinos may also be feasible, 
however - as Eq. (3) indicates - a careful accounting of matter effects for $\anumu$ as well as $\numu$ fluxes 
with consideration of mass hierarchy is required.

\vspace{+5pt}
\section*{Acknowledgments} \vspace{-8pt}  This work was supported by the United States Department of Energy under grant DE-SC0007866.

%%%%%%%%%%%%%%%%%%%%%%%%%%%%%%%%%%%%%%%%%%%%%%%%%%%%%%%%%%%%%%%%%%%%%%%

\end{document}